\documentclass[preprint]{revtex4-1}
\usepackage{graphicx}
\usepackage{dcolumn}
\usepackage{bm}
\usepackage{color}
\usepackage{soul}

\usepackage{xr}
\makeatletter
\newcommand*{\addFileDependency}[1]{
  \typeout{(#1)}
  \@addtofilelist{#1}
  \IfFileExists{#1}{}{\typeout{No file #1.}}
}
\makeatother
 
\newcommand*{\myexternaldocument}[1]{%
    \externaldocument{#1}%
    \addFileDependency{#1.tex}%
    \addFileDependency{#1.aux}%
}
\myexternaldocument{Suppl_info}

\usepackage{amsmath}
\usepackage{amssymb}

\newcommand{\nc}{\newcommand}
\nc{\ta}{\tilde{a}}
\nc{\I}{$I$}
\nc{\II}{$II$}
\nc{\III}{$III$}
\nc{\nn}{\nonumber}
\def\e{\mathcal{E}}

\begin{document}

\bibliographystyle{naturemag}

\title{
Reply to comment on "Thermal effects - an alternative mechanism for plasmon-assisted photocatalysis"}

\author{Yonatan Dubi,$^{1,4}$ Ieng Wai Un,$^{2,3}$, Yonatan Sivan$^{2,4\ast}$ 
\\
\normalsize{$^{1}$Department of Chemistry, Ben-Gurion University, Israel }\\
\normalsize{$^{2}$School of Electrical and Computer Engineering, Ben-Gurion University of the Negev, Israel}\\
\normalsize{$^{3}$Joan and Irwin Jacobs TIX Institute, National Tsing Hua University, Taiwan}\\
\normalsize{$^{4}$ Ilse Katz Center for Nanoscale Science and Technology, Ben-Gurion University, Israel}\\
\normalsize{$^\ast$To whom correspondence should be addressed; E-mail: jdubi@bgu.ac.il}
}

\date{\today}

\maketitle

In his Comment to our paper "Thermal effects -- an alternative mechanism for plasmon-assisted photocatalysis"~\cite{dubi2020thermal}, Jain correctly points out that using an Arrhenius fit to the reaction rate is not enough to distinguish thermal from non-thermal effects. The reason is that the Arrhenius form contains only the ratio $\e_a/(k_B T)$, where $\e_a$ is the reaction activation energy, and $T$ is the catalysts (average) temperature ($k_B$ is the Boltzmann constant). Illumination causes an increase of the temperature, but some claim that it may also reduce the activation energy (due to the generation of nonthermal, "hot",  carriers). It might also do both,  simultaneously. Thus, there is a continuum of possible fits to the reaction rate (as a function of temperature, say), ranging from the limit where only the activation energy changes, to the limit where the activation energy remains constant and only the temperature changes (as done in our manuscript). Mathematically, this implies that if $T = T_{dark} + a I_{inc}$ (where $T_{dark}$ is the ambient temperature, $I_{inc}$ is the incident illumination intensity and $a$ is the photothermal conversion coefficient), then $a$ can range from a maximal value all the way down to zero, and still yielding a good fit to the data. 

Jain is in fact correct, which is why, for example, we used the term ``alternative explanation'' in our title, to demonstrate the caution practiced in our manuscript. Furthermore, {\sl this precise point was raised in our recent publication} (Y. Sivan, J. H. Baraban, and Y. Dubi, OSA Continuum 2020, 3, 483-497)~\cite{R2R}, Section 4.8. In Fig.~1 below (taken from~\cite{R2R}) we show fits to the data of Zhou et al.~\cite{Halas_Science_2018} (Ref. [48] in~\cite{dubi2020thermal}), where the photo-thermal conversion coefficient ranges from $a = 0$ to $a = 180$ K/W cm$^{-2}$ (left panels). As can be clearly seen, the fits are excellent (in fact indistinguishable). On the right panels of the figure we plot the resulting dependence of activation energy on illumination intensity. Clearly, if one assumes no heating ($a = 0$) then the activation energy strongly depends on intensity, while it is essentially constant if $a$ is maximal.  

Why, then, in our paper~\cite{dubi2020thermal} we seemingly discuss only the limit of constant activation energy? This is discussed in Section II of Ref.~\cite{dubi2020thermal}. The key point is that, as our theoretical work demonstrated~\cite{Dubi-Sivan}, two points seem to contradict the idea that ``hot'' electrons somehow contribute to the reaction rates in the papers we discuss in Ref.~\cite{dubi2020thermal}. The first is that only a tiny fraction of the illumination power ($\sim 10^{-7}-10^{-10}$) actually goes to generating hot electrons, and the second is that the {\sl number} of ``hot'' electrons nevertheless increases by many orders of magnitude, but reaction rates only rise by a moderate 1-2 orders of magnitude~\cite{Dubi-Sivan-Faraday}. 

Moreover, in Ref.~\cite{dubi2020thermal} we evaluate the photothermal conversion coefficient independently from either fitting to the data (Fig.~5) or from any direct calculation (Section~II of the SI to Ref.~\cite{Dubi-Sivan}), and find very similar values to those found in the constant activation energy fit. So, by virtue of Occam's razor, the explanation of thermal effects, which is simpler {\sl and} corroborated by independent quantitative calculations, is far more likely to be the correct one (compared with the speculative claim for ``hot'' electron action which is not backed up by any sort of theory).  

Finally, in the last paragraph of his comment, Jain reminds the readers of the importance of temperature gradients. The importance of thermal gradients was already discussed in great detail in both Section IV of Ref.~\cite{dubi2020thermal}, and in our Ref.~\cite{R2R} 
In this context, we point that Jain includes Ref.~\cite{Halas_Science_2018} in the list of ``practitioners acknowledging the importance of temperature gradients'', while (as we show in~\cite{anti-Halas_comment} and~\cite{R2R}), the authors of Ref.~\cite{Halas_Science_2018} may have measured them incorrectly, and do not  acknowledge their importance.  

\begin{figure}[ht]
\centering
\includegraphics[width=0.9\textwidth]{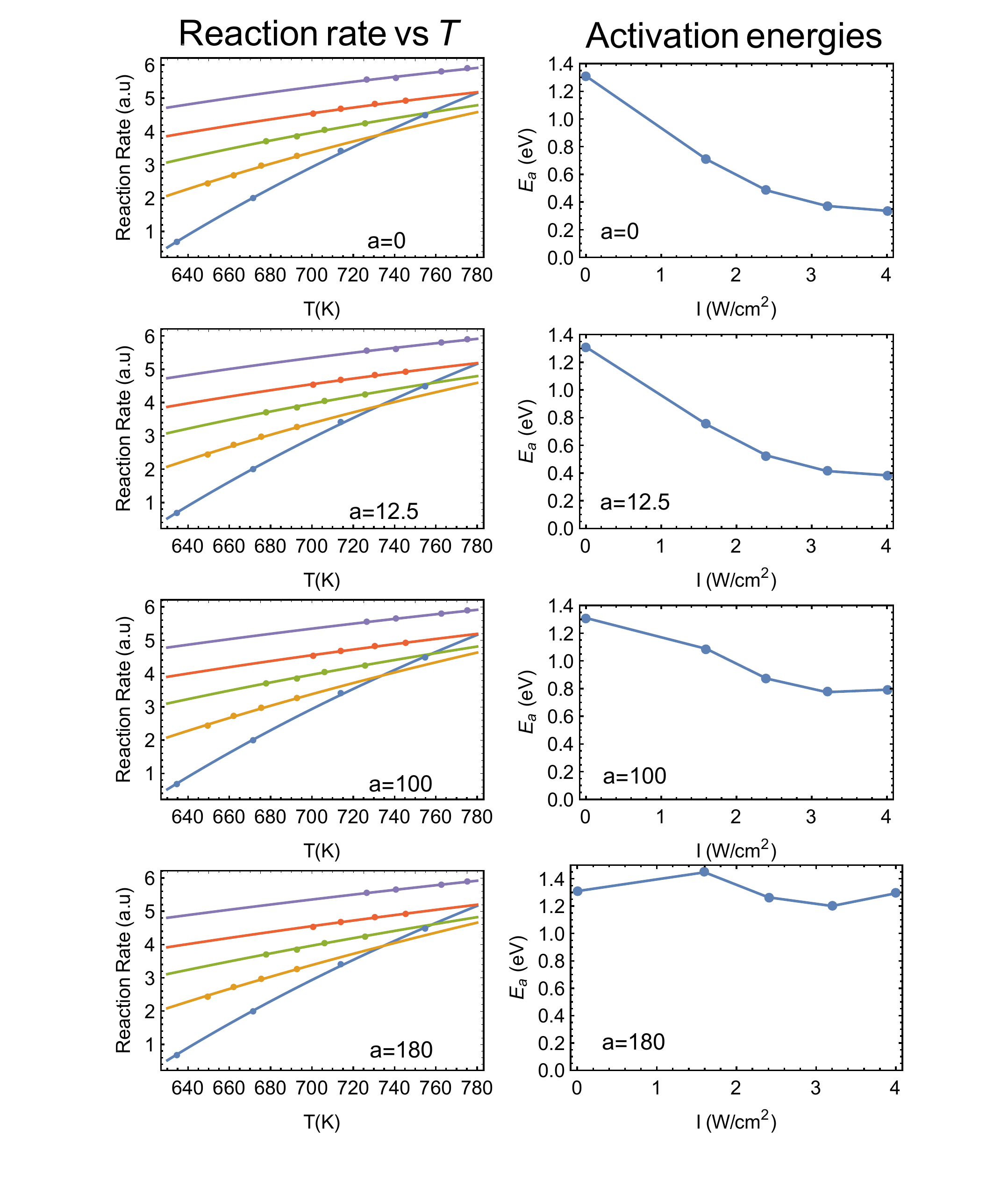}
\caption{Left panels: reaction rate as a function (inverse) temperature, points are data from Ref.~\cite{Halas_Science_2018} (Ref.~[48] in Ref.~\cite{dubi2020thermal}. The solid lines are fits to an Arrhenius form with varying values of $a$. Right panels: the resulting activation energy as a function of intensity, going from a strongly intensity-dependent activation energy (this is what is plotted in Fig.~2C of Ref.~\cite{Halas_Science_2018}), all the way to an essentially intensity-independent activation energy for $a=180$ K/W cm$^{-2}$. This figure is taken from Ref.~\cite{R2R}.}  \label{fig:fitting_a}
\end{figure}

\end{document}